\documentclass[twocolumn,prd,unsortedaddress,superscriptaddress,showpacs,a4paper,nofootinbib]{revtex4}

\usepackage{graphics}
\def\MeV{\mbox{ MeV}}
\def\GeV{\mbox{ GeV}}

\def\br{\begin{eqnarray}}
\def\er{\end{eqnarray}}
\def\be{\begin{equation}}
\def\ee{\end{equation}}

\usepackage{graphicx}
\usepackage{dcolumn}
\usepackage{bm}
\usepackage{latexsym} 
\usepackage{dsfont}
\usepackage{slashed}
\usepackage{amsmath}

\begin{document}

\title{\sc Nonleptonic $B$ meson decays in collinear pQCD at twist-3: Effects of dynamical
masses of gluons and quarks}

\author{C.~M.~Zanetti} 
\email{carina@if.usp.br}
\affiliation{Instituto de F\'{\i}sica, Universidade de S\~{a}o Paulo, 
C.P. 66318, 05389-970 S\~{a}o Paulo, SP, Brazil}

\author{A.~A.~Natale}
\email{natale@ift.unesp.br}
\affiliation{Instituto    de    F\'{\i}sica    Te\'orica,   UNESP    -
  Univ.  Estadual Paulista,  Rua Dr.  Bento  T. Ferraz,  271 -  Bl.II,
  01140-070, S\~ao Paulo - SP, Brazil}

\begin{abstract}
We compute the amplitudes  for non-leptonic annihilation decays of $B$
mesons into two particles within  the pQCD collinear approach. The end
point divergences  are regulated with  the help of an  infrared finite
gluon propagator  characterized by a  non-perturbative dynamical gluon
mass consistent  with recent  lattice simulations. The  divergences at
twist-3 are regulated by a dynamical quark mass. Our results fit quite
well   the   existent   data   of  $B^0\to   D_s^-K^+$   and   $B^0\to
D_s^{-*}\bar{K}^+$   for  the  expected   range  of   dynamical  gluon
masses. We also make predictions  for the rare decays $\bar{B}^0\to K^-K^+$,
$\bar{B}^0_s\to    \pi^-\pi^+,\pi^0\pi^0$,   $B^+\to   D_s^{(*)+}\bar{K}^0$,
$B^0\to D_s^{\pm(*)} K^\mp$ and $B_s^0\to D^{\pm(*)} \pi^\mp, D^0\pi^0$.
\end{abstract}

\pacs{  {12.38.Bx}, {12.38.Aw}, {12.38.Lg}, {13.25.Hw} }

\keywords{B meson; hadronic decays; infrared cutoff.}

\maketitle

%

%

\section{Introduction}

Perturbative QCD  has been largely applied  in the study  of $B$ meson
decays,   specially  in   non-leptonic  decays   which   requires  the
calculation of  hadronic matrix elements.  In such  processes the long
distance  interactions  between the  initial  and  final state  mesons
impose a  difficult obstacle in amplitudes calculations.   In the pQCD
approach the calculation of the matrix elements can be performed since
the perturbative dynamics can  be factorized from the non-perturbative
dynamics  by  applying  the  hard  exclusive  scattering  Brodsky  and
Lepage's  approach  \cite{brodsky2,brodsky}.   In this  procedure  the
non-perturbative  contribution  in  exclusive hadronic  processes  are
contained in  the wave  functions that describe  the hadrons  as bound
states  of quarks.  The  employment of  pQCD is  justified due  to the
large scale involved  in the strong interactions, since  the energy of
gluons exchanged in  such processes are of the order  of the $B$ meson
mass \cite{brodsky1}.

There  are  some  problems  in  this  approach. One  of  them  is  the
occurrence of  soft divergences for  the vanishing gluon  momentum, as
well  as collinear  divergences  when  the momentum  of  the gluon  is
parallel to  the momentum  of a massless  quarks (since the  masses of
quarks $u,d,s$ are usually neglected).  Double logarithmic divergences
also occur  with both soft and collinear  divergences.  Usually, these
divergences  can be  absorbed into  the meson  wave functions  and the
factorization is  preserved. However  in the amplitude  calculation of
$B$ meson decays may appear endpoint divergences that can not absorbed
into the wave functions.  This happens when there are non-factorisable
contributions related to interactions with the spectator quark and for
annihilation interactions, respectively at twist-3 and twist-2.  These
contributions, in general, are  power suppressed and are negligible in
front of the factorisable  contributions.  However, it is important to
have a  reliable calculation of  the annihilation diagrams  since they
can generate strong  phases that are relevant in  the determination of
the CP  violation parameters, as well  as in the  evaluation of decays
that occurs through pure annihilation processes.

In order to deal with  these divergences several approaches have being
proposed in recent years.  In  pQCD approach by H.~N.~Li {\it et.  al}
\cite{pqcd,pqcd1,pqcd3,Lu:2000em,pqcd2}   a  $k_T$   factorization  is
performed, where the transversal momentum of the quarks are taken into
account.  The  large logarithms are then resumed  resulting in Sudakov
factors  that suppress the  soft dynamics.   In the  QCD factorization
approach      (QCDF)     proposed      by      Beneke     et.       al
\cite{beneke,beneke1,beneke2,beneke3}, the perturbative QCD is used at
some extent.  The authors argue that the transition form factors, that
enter  in the  calculation of  $B$  meson decays,  cannot be  obtained
through perturbative  methods because  they are dominated  by ``soft''
interactions, and the Sudakov factors are not enough to suppress them.
On  the  other  hand,  it  is  considered  that  the  non-factorisable
contributions are  dominated by  hard gluon exchanges.   Therefore, in
QCDF the terms  that are dependent on the  transition form factors are
parameterized as in the  naive factorization, and the non-factorisable
contributions   are  perturbatively   calculated   in  the   collinear
approximation.  The endpoint divergences in non-factorisable diagrams,
are   treated   with    a   phenomenological   and   model   dependent
parameterization as  in the following equation  \be \int \frac{dx}{x}=
\ln \frac{m_B}{\Lambda_h}(1+\rho e^{\imath  \varphi}) , \,\,\,\, 0\leq
\rho\leq 1.
\label{eqcut}
\ee Although this computation is an \textsl{ad hoc} one it allow us to
use   this   approach   to   estimate   the   contributions   of   the
non-factorisable diagrams.

One third possibility that has been successfully applied several times
in order to  deal with the divergences in  the collinear pQCD approach
is    the     use    of     an    IR    finite     gluon    propagator
\cite{yang,yang1,yang2,zanetti,yang3,yang4},   which   appears  as   a
non-perturbative  solution of  the  gluonic Schwinger-Dyson  equations
(SDE) \cite{cornwall} and has  been confirmed by lattice simulation of
pure  QCD \cite{abp}.   This gluon  propagator is  characterized  by a
dynamical gluon  mass, which naturally provides  a self-consistent and
model independent  calculation.  This  approach works well  at leading
twist \cite{zanetti},  but at  twist-3 collinear divergences  may also
appear.    These    divergences   occur   in    the   massless   quark
approximation.  Of course  such divergences  may be  cured introducing
current quark  masses, which lead to an  unnatural strong contribution
from  the fermion  propagators  of the  light  $u$ and  $d$ quarks  at
twist-3. Moreover  for consistency with  the gluon dressing  we cannot
neglect  the fermion  dressing  that gives  dynamical  masses of  order
$\Lambda_{QCD}$ to the light quarks.

In  this  work,  we  will  study the  regulation  of  the  divergences
occurring in  the collinear factorization  of pQCD.  We focus  on pure
annihilation  $B$  meson  decays,  in  which  the  dependence  on  the
regulation will be more evident.  We calculate the branching ratios of
several pure annihilation channels of $B^0_d$, $B^+$ and $B^0_d$.  Our
main  goal is  to revisit  and  improve our  previous calculations  at
twist-2  \cite{zanetti},   using  an  infrared   gluon  propagator  in
agreement with the most  recent QCD lattice simulations and performing
the  twist-3  calculation.   In  order  to  deal  with  the  collinear
divergences that appear at twist-3 we will use the dynamical masses of
the light  quarks. This approach  is mandatory if we  want consistency
with the use of dynamical  gluon masses.  We study the following decay
channels       into      pseudo-scalar       particles:      $B_s^0\to
\pi^+\pi^-,\,D^{\pm}\pi^{\mp}$,  $B_d^0\to  K^+K^-,\,D_s^{\pm}K^{\mp}$
and $B^+\to D_s^+\bar{K}^0$; and  also the decay into a pseudo-scalar
and   a  vector   particle,   $B_d^0\to  D_s^{*\mp}K^\pm$,   $B_d^0\to
D_s^{*\mp}K^\pm$ and $B^+\to D_s^{*+}\bar{K}^0$.  We will then compare
our results  with the available data.   We have to  mention that these
decays  are  quite rare,  and  most of  them  are  beyond the  present
experimental limits.  It is expected however that they can be detected
with the advent of LHCb (LHC, CERN), or even earlier by the experiment
CDF (Tevatron, Fermilab).

The distribution of  the paper is the following: In  Sec. II we present
the  basic  expressions  of  the  pQCD  approach  to  calculate  the
amplitudes of $B$  meson decays.  In Sec.  III  we show the annihilation
amplitudes for  the different $B$ decays.   In Sec.  IV  we discuss few
aspects of dynamical mass generation and the
regulation  of  the  divergences.  Section V  contains  our  numerical
results and Section VI is devoted to our conclusions.

\section{Perturbative QCD approach}

Weak decays of $B$ mesons are described by an effective Hamiltonian at
a renormalization  scale $\mu\ll M_W$.  This  effective Hamiltonian is
formally  obtained   using  the  Operator   Product  Expansion  (OPE),
describing effective interactions through four-quark operators, and is
given by \cite{buras}:
\begin{equation}\label{ham}
  {\mathcal H}_{\mathrm{eff}}=\frac{G_F}{\sqrt{2}}V_{\mathrm{CKM}}\sum_iC_i(\mu)\,Q_i(\mu),
\end{equation}
where   $G_F$   is   the   Fermi  constant,   $V_{\mathrm{CKM}}$   are
Cabibbo-Kobayashi-Maskawa factors, $Q_i$ are the four-quarks operators
contributing to the decay and $C_i(\mu)$ are the Wilson coefficients .
The amplitude of the decay $B\to f$ is then obtained from:
\begin{eqnarray}\label{amp}
{\mathcal  A}(B\to f)  & = & \langle  f\vert{\mathcal
H}_{\mathrm{eff}}\vert B\rangle   \nonumber\\&=&\frac{G_F}{\sqrt{2}}\sum_i V_{\mathrm{CKM}}^i C_i(\mu)\langle   Q_i \rangle\,,
\end{eqnarray}
\noindent  where  $\langle  Q_i  \rangle=\langle  f\vert  Q_{i}  \vert
B\rangle$  are the hadronic  matrix elements  between the  initial and
final states.

To compute the  amplitudes we use the pQCD  formalism in the collinear
approximation, where the hadronic matrix elements are obtained through
a  convolution of  the hard  scattering kernel  and  the distribution
amplitudes of  the mesons involved in  the process.  In the  case of a
two-body non-leptonic decay $B$  $\to$ $M_1M_2$, the matrix element of
the operator $Q_i$ is given by:
\begin{eqnarray}\label{amppqcd}
 \langle   Q_i \rangle =\int \mathrm{d}{x}\,\mathrm{d}{y}\,\mathrm{d}{z}\, T_i(x,y,z)\,
\Phi_{M_1}(x)\,\Phi_{M_2}(y)\,\Phi_B(z),
\end{eqnarray}
where $x,y,z$  are the momentum  fractions, $\Phi_{\mathrm{M}}(x)$ are
the  light-cone   distribution  amplitudes  for   the  quark-antiquark
sta\-tes of  the mesons, which  are non-perturbative functions  of the
momentum fraction carried by the partons; $T_i$ is the hard scattering
kernel  that  can  be  perturbatively  computed  as  function  of  the
light-cone momenta  of collinear partons. The branching  ratio is then
calculated from
\begin{equation}
{\mathcal B}(B\to M_1M_2)=\frac{\tau_B p_c}{8\pi m_B^2} 
\vert\mathcal{A} (B\to M_1M_2)\vert^2,
\end{equation}
where $\tau_B$ is the $B$ meson lifetime, $p_c$ is the momentum of the
final state particles with masses $m_1$  and $m_2$ in the $B$ meson rest
frame,
\begin{equation}
p_c=\frac{1}{2m_B}\sqrt{(m_B^2-(m_1+m_2)^2)(m_B^2-(m_1-m_2)^2))}.
\end{equation}

As usual, it is used the light-cone distribution amplitudes decomposed
in terms of spin structure. The decomposition of the wave functions of
the  $B$ mesons  are expressed  in terms  of two  Lorentz  scalar wave
functions:
 \begin{eqnarray}
    \Phi_{B,(\alpha\beta)}(x)&=&\frac{i}{\sqrt{2 N_c}}[(\slashed{P}\gamma_5)_{\alpha\beta}+m_{B}(\gamma_5)_{\alpha\beta}]\times\nonumber\\& \times &\left(\phi_{B}(x)+\frac{\slashed{n}}{\sqrt{2}}\bar{\phi}_{B}(x)\right)\,,
  \end{eqnarray}
  where the normalization of the scalar functions are the following:
  \be
  \int_0^1dx\,\phi_B(x)=\frac{f_B}{2\sqrt{6}}\,;\qquad \int_0^1dx\,\bar{\phi}_B(x)=0\,.\label{normbwave}
  \ee  
  However,  as  it was  shown  in Refs.~\cite{Lu:2002ny,Kurimoto},  the
  contribution from the first  wave function is dominant, therefore in
  our calculations we will consider only the first term of the Lorentz
  structure.

For  the $D$ pseudoscalar  mesons we  use the  following decomposition
\cite{Kurimoto}:
 \begin{equation}
    \Phi_{D(\alpha\beta)}(x)=\frac{i}{\sqrt{2 N_c}}\left[(\gamma_5\slashed{P})_{\alpha\beta}+m_{D}(\gamma_5)_{\alpha\beta}\right]\phi_{D}(x)\,.
  \end{equation}

The light pseudo-scalars mesons are represented by the following wave function 
\begin{eqnarray}
\Phi_{K,\pi(\alpha\beta)}&=&\frac{i}{\sqrt{2N_c}}[\gamma_5\slashed{P}\phi_A(x)+m_{0P} \gamma_5\phi_P(x)+ \nonumber\\ &&+m_{0P}\gamma_5(\slashed{n}\slashed{v}-1)\phi_T(x)]_{\alpha\beta},\label{lightwave}
\end{eqnarray}
where the wave  function $\phi_A(x)$ is the twist-2  wave function and
$\phi_P(x),\phi_T(x)$     are    twist-3    wave     functions,    and
$m_{0P}=M_P^2/(m_{q1}+m_{q2})$   ($m_{0\pi}=1.5\GeV,\,m_{0K}=1.6\GeV$).
The vectors  $n,v$ are light-like vectors:  for a meson  moving in the
direction  $P\propto(1,0,{\bf0}_T)$,   the  vector  $n=(1,0,{\bf0}_T)$
defines  the  direction  of  motion   of  the  meson,  and  we  define
$v=(0,1,{\bf0}_T)$.  For  a meson  moving with momentum  defined along
the  $v$  direction,   the  third  term  of  Eq.~(\ref{lightwave}),   we  have  that
$\slashed{n}\slashed{v}$ is replaced by $\slashed{v}\slashed{n}$.

For the heavy vector  mesons ($D_{(s)}^{*\pm}$), only the longitudinal
part of  the wave  function contributes to  the amplitude, due  to the
conservation of angular momentum.  We use the following expression:
\begin{equation}
\Phi_{D^*,\alpha\beta}(x)=\frac{i}{\sqrt{2 N_c}}\slashed{\epsilon}[(\slashed{P}\gamma_5)_{\alpha\beta}+m_{D^*}(\gamma_5)_{\alpha\beta}]\phi_{D^*} (x),
\end{equation}
where    $\epsilon=\frac{1}{\sqrt{2}r}(1,-r^2,\vec{0}_T)$    is    the
longitudinal polarization of the vector meson, with $r=m_D/m_B$.

\begin{figure}[pt]
\begin{center}
  \includegraphics[width=8cm]{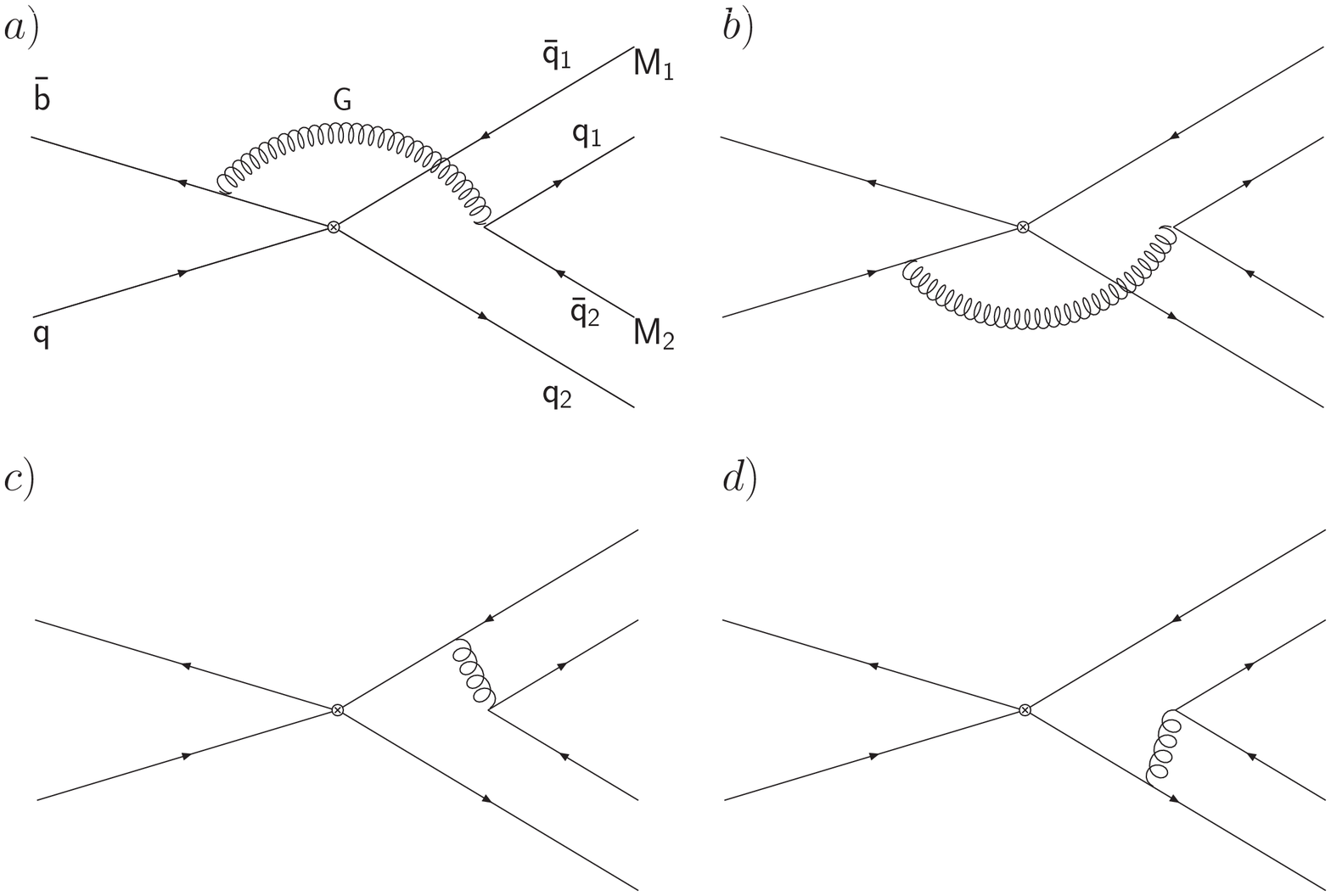}
\caption{   Feynman  diagrams   contributing  to   the   amplitude  of
  annihilation  non-leptonic   decay  channels.  The  non-factorisable
  diagrams  are  a) and  b),  and the  factorisable  ones  are c)  and
  d).}\label{figure1}
\end{center}
\end{figure}

\section{Annihilation Amplitudes}

Annihilation decays  are processes that  occur at $\alpha_s$  order in
perturbative QCD.  A typical annihilation  decay of a B meson is shown
in  Fig.~\ref{figure1}, where the  bottom quark  decays through  the W
exchange  or via  a  penguin  process and  a  quark-antiquark pair  is
created in  the final  state through  a gluon emission  by any  of the
quarks   involved   in  the   process.    The   channels  with   these
characteristics that we shall analyze are the following:
\begin{itemize}
\item  Charmless channels : 

$\bar{B}^0_s\to\pi^+\pi^-,\pi^0\pi^0$; $\bar{B}^0_d\to
  K^+K^-$;
\item Charmed  channels: 

$B^0\rightarrow  D_s^{\pm(*)} K^{\mp}$; $B^+\rightarrow  D_s^{+(*)} \bar{K}^{0}$;

$B_s^0  \rightarrow D^{\pm(*)}\pi^{\mp},D^0\pi^0,\bar{D}^0\pi^0$.

\end{itemize}
In  the   following  we  present  the  analytical   formulas  for  the
computation of the amplitudes for the non-leptonic decays of B mesons.

\subsection{Charmless channels}

We  present the  amplitudes  for the  diagrams of  Fig.~\ref{figure1},
obtained  with the  help  of Eq.~(\ref{amppqcd}).   The diagrams  that
actually  contribute for  the amplitude  of charmless  decays  are the
non-factorisable   diagrams    shown   in   Figs.~\ref{figure1}a   and
\ref{figure1}b,  since the  contributions from  the  factorisable ones
(Figs.~\ref{figure1}c  and  \ref{figure1}d)  cancel among  themselves.
The full amplitude for the decays $\bar{B}\to\pi\pi,KK$ is given by
\begin{eqnarray}\label{ampcharmless}
{\mathcal A}(\bar{B}\to M_1M_2)& = & \frac{G_F}{\sqrt{2}}\,f_Bf_{M_1}f_{M_2}( V_{ub}\,V_{uq}^*\,{\mathcal A}_{\mathrm{Tree}}\nonumber\\&& - V_{tb}\,V_{tq}^*\,{\mathcal A}_{\mathrm{Penguin}}),
\end{eqnarray}
\noindent  with $q=d$  and $q=s$  for the  decays of  the  $B_d^0$ and
$B_s^0$  mesons, respectively.   In particular  the amplitude  for the
decay  mode  $\bar B_s^0\to\pi^0\pi^0$  relates  to  the one  of  the  mode
$\pi^+\pi^-$                       as                       ${\mathcal
  A}(\bar B_s^0\to\pi^0\pi^0)=\frac{1}{\sqrt{2}}{\mathcal
  A}(\bar B_s^0\to\pi^+\pi^-)$.  

We  consider  the   $B$  meson  at  the  rest   frame,  with  momentum
$P_B=\frac{m_B}{\sqrt{2}}(1,1,{\bf0}_T)$  in  light-cone  coordinates
$(p^+,$  $p^-, {\bf  p}_T)$, with $p^\pm={1\over\sqrt{2}}(p^0\pm  p^3)$ and
${\bf   p}_T=(p^1,p^2)$  (${\bf   p}_T=\bf{0}_T$   in  the   collinear
approximation).  The  mesons $M_1$ and  $M_2$ have momenta  defined as
$P_1=\frac{m_B}{\sqrt{2}}(0,1,{\bf0}_T)\propto          v$         and
$P_2=\frac{m_B}{\sqrt{2}}(1,0,{\bf0}_T)\propto        n$,        hence
$P_B=P_1+P_2$. We also define the  momenta of the quarks in the mesons
$B,M_1,M_2$       respectively       as       $q=(zP_B^+,0,{\bf0}_T)$,
$q_1=(0,xP_1^-,{\bf0}_T)$, $q_2=(yP_2^+,0,{\bf0}_T)$  (the momentum of
the antiquarks are given by $\bar{q}=P-q$).

The tree level and penguin amplitudes are 
\begin{equation}
{\mathcal A}_{\mathrm{Tree}}=C_2\,{\mathcal A}_{1}\,,
\end{equation}
\begin{equation}
{\mathcal A}_{\mathrm{Penguin}}=\left( 2\,C_4 + \frac{C_{10}}{2}\right)\,{\mathcal A}_{1}+\left(2\,C_6 + \frac{C_8}{2}\right)\,{\mathcal A}_{2}.
\end{equation}

The  functions   ${\mathcal  A}_1$   and  ${\mathcal  A}_2$   are  the
contributions  from  operators  of  the type  $(V-A)\otimes(V-A)$  and
$(V-A)\otimes(V+A)$, respectively, and are given by

\begin{eqnarray}
&&{\mathcal A}_{1}=\frac{32 \pi m_B^4 C_F}{\sqrt{2 N_C}} \int_0^1 dx\,dy\,dz\,{\phi_B}(z) \biggl\{  h_b(x,y,z)\times\nonumber\\&\times&\biggr[
   {-\bar{x}\phi_A}(x) {\phi_A}(y)+\mu_{0P}^2\biggl(-4 {\phi_P}(x)
   {\phi_P}(y)+\left(\bar{z}-y\right) \times\nonumber\\&\times&({\phi_P}(x)+{\phi_T}(x) ({\phi_P}(y)-{\phi_T}(y))+x({\phi_P}(x)-{\phi_T}(x))\times\nonumber\\&\times&({\phi_P}(y)+{\phi_T}(y))\biggr)\biggr] +
h_q(x,y,z)\biggl[(y-z)\phi_A(x) \phi_A(y)+\nonumber\\ &  +&\mu_{0P}^2\biggl(\bar{x} ({\phi_P}(x)+{\phi_T}(x)) ({\phi_P}(y)-{\phi
   T}(y))+\nonumber\\ &  + &(y-z) ({\phi_P}(x)-{\phi_T}(x)) ({\phi_P}(y)+{\phi_T}(y))\biggr)\biggr]\biggr\},
\end{eqnarray}
and
\begin{eqnarray}
& & {\mathcal A}_{2}=\frac{32\pi m_B^4 C_F}{\sqrt{2 N_C}}\int_0^1 dx\,dy\,dz\, {\phi_B}(z) \biggl\{h_b(x,y,z)\times\nonumber\\&\times&\biggl[\left(-y+\bar{z}-1\right){\phi_A}(x) {\phi_A}(y)+\mu_{0P}^2\biggl(-4 {\phi_P}(x)
   {\phi_P}(y)+\nonumber\\& +&x ({\phi_P}(x)+{\phi_T}(x))
   ({\phi_P}(y)-{\phi_T}(y))+\left(\bar{z}-y\right)
   ({\phi_P}(x)+\nonumber\\&-&{\phi_T}(x))({\phi_P}(y)+{\phi_T}(y))\biggr)
  \biggr]+h_q(x,y,z)\biggl[\bar{x}   {\phi_A}(x) {\phi_A}(y)+\nonumber\\ & &+{\mu_{0P}^2}\biggl((y-z) ({\phi)_P}(x)+{\phi_T}(x)) ({\phi_P}(y)-{\phi_T}(y))+\nonumber\\ & &+\bar{x} ({\phi_P}(x)-{\phi_T}(x)) ({\phi_P}(y)+{\phi_T}(y))\biggr) \biggr]\biggr\}.
\end{eqnarray}
Where $\mu_{0P}=m_{0P}/m_B$, and the functions $h(x,y,z)$ are 
\begin{eqnarray}
h_b(x,y,z)&=&\alpha(\mu)D_b(x,y,z)D_g(x,y)\nonumber\\
h_q(x,y,z)&=&\alpha(\mu_h)D_q(x,y,z)D_g(x,y),
\end{eqnarray}
and the quarks and gluon perturbative propagators are defined as:
\begin{eqnarray}
D_b^{-1}(x,y,z)&=&m_B^2x(\bar{z}-y)-m_b^2\nonumber\\
D_q^{-1}(x,y,z)&=&m_B^2\bar{x}(y-z)-m_q^2\nonumber\\
D_g^{-1}(x,y)&=&(k_g^2)^{-1}=m_B^2y\bar{x}
\end{eqnarray}
The  scales  adopted  in   the  calculations  are  $\mu=m_b$  for  the
contribution from  the diagram shown in Fig.  \ref{figure1}a where the
gluon is  emitted from the $b$  quark, and $\mu_h=\sqrt{\Lambda_hm_b}$
for the contribution shown in  Fig.  \ref{figure1}b where the gluon is
emitted   from   the  spectator   quark,   with  $\Lambda_h=500$   MeV
\cite{beneke}.    When   the   effective   scale is  $\mu<m_b$   we   use
$\Lambda_{QCD} = 225$ MeV and when $\mu = m_b$ we use $\Lambda_{QCD} =
300$  MeV, and  the  difference is  to  match the  values  of the
coupling constant due to different quark thresholds.

\subsection{Charmed channels}

The charmed  decay channels occur through $W$  exchange processes, and
there is  no contribution from  penguin diagrams.  The  full amplitude
for    the   general    processes    $B\to   D^{\mp}M^{\pm}$    (with
$B=B_d,B_s^0,B^+$;    $D,M={D_{(s)}^{\pm(*)},K,\pi}$)    will    be
calculated from
\begin{eqnarray}                 {\mathcal                  A}(B\to
  D^{-}M^{+})&=&\frac{G_F}{\sqrt{2}}V_{uq}V_{cb}^*\,{\mathcal{M}}({B\to D^{-} M^+})\,,\nonumber\\
 {\mathcal                  A}(B\to
  D^{+}M^{-})&=&\frac{G_F}{\sqrt{2}}V_{cq}V_{ub}^*\,{\mathcal{M}}({B\to D^{+} M^-})\,,\label{ABDM}
\end{eqnarray}
where $q=d$ for the decays of $B^0$ and $B^+$, and $q=s$ for $B_s^0$.

We  consider  the   $B$  meson  at  the  rest   frame,  with  momentum
$P_B=\frac{m_B}{\sqrt{2}}(1,1,{\bf0}_T)$.  The  light mesons $M=\pi,K$
have                 momenta                 defined                as
$P_M=\frac{m_B}{\sqrt{2}}(0,1-r^2,{\bf0}_T)\propto  v$,  and  for  the
charmed           mesons           the           momentum           is
$P_D=\frac{m_B}{\sqrt{2}}(1,r^2,{\bf0}_T)$,  hence $P_B=P_M+P_D$.  The
momentum of the quarks are defined analogously to the charmless case.

The  functions ${\mathcal{M}}$ in  the previous  Eqs.~(\ref{ABDM}) are
the   sum   of   factorisable  and   non-factorisable   contributions,
${\mathcal{M}}={\mathcal{M}}_{\mathrm{fac}}+{\mathcal{M}}_{\mathrm{nfac}}$.
The analytical  formulas for these  contributions for the  decays with
two pseudo-scalar particles in the final state, are the following:
\begin{itemize}
\item ${B^0 \to D^-M^+}$:
\end{itemize}

\begin{eqnarray}\label{BDK1}
& &  {\mathcal{M}}_{\mathrm{fac}}({{B^0 \to D^-M^+}})  =  \frac{8\pi m_B^4C_F}{\sqrt{2N_C}}\int_0^1   {dx}\,{dy}\,{dz} \nonumber\\ &\times&\phi_B(z)\phi_{D}(y)\biggr\{-\biggr((1-r^2)y\phi_A(x)-2r\mu_{0P}(1-r^2+y)\phi_P(x)\biggl)\times\nonumber\\ &\times&h_{l}(x,y,z)-
\biggr((1-r^2)(r^2-(1-r^2)\bar{x})\phi_A(x) -r\mu_{0P}(3-r^2+\nonumber\\&-&2(1-r^2)x)\phi_P(x)-(1-r^2)(1-2x)\phi_T(x) \biggl)h_{c}(x,y,z)
\biggl\}\,,
\end{eqnarray}
\begin{eqnarray}\label{BDK2}
& &  {\mathcal{M}}_{\mathrm{nfac}}({{B^0 \to D^-M^+}})   =   \frac{32\pi m_B^4C_F}{\sqrt{2N_C}}\int_0^1   {dx}\,{dy}\,{dz}\phi_B(z)\phi_{D}(y)\times\nonumber\\ &\times&\biggr\{-\biggr((1-r^2)((1+r^2)(y-\bar{z})-1)\phi_A(x)-r\mu_{0P}\bigr((2+y+z+\nonumber\\ &+&(1-r^2)\bar{x})\phi_P(x)+(-y+z+(1-r^2)\bar{x})\phi_T(x)\bigl)\biggl)h_b(x,y,z)+\nonumber\\ &+&\biggr((1-r^2)((y-z)r^2+(1-r^2)\bar{x})\phi_A(x) +r\mu_{0P}\bigr((y-z+\nonumber\\ &+&(1-r^2)\bar{x})\phi_P(x)+(y-z-(1-r^2)\bar{x})\phi_T(x)\bigl) \biggl)h_{q}(x,y,z)\biggl\}\,.
\end{eqnarray}

\begin{itemize}
\item ${B^0 \to D^+M^-}$:
\end{itemize}

\begin{equation}\label{BDK3}
{\mathcal{M}}_{\mathrm{fac}}({{B^0 \to D^+M^-}})=-{\mathcal{M}}_{fac}({{B^0 \to D^-M^+}})
\end{equation}

\begin{eqnarray}\label{BDK4}
&& {\mathcal{M}}_{\mathrm{nfac}}({{B^0 \to D^+M^-}})  = \frac{32\pi m_B^4C_F}{\sqrt{2N_C}}\int_0^1   {dx}\,{dy}\,{dz} \times\nonumber\\ &\times&\phi_B(z)\phi_{D}(y)\biggr\{-\biggr((1-r^2)((1-r^2)\bar{x}+(y-\bar{z})r^2)\phi_A(x)+\nonumber\\ &-&r\mu_{0P}\bigr((2+y+z+(1-r^2)\bar{x})\phi_P(x)+((1-r^2)\bar{x}+\nonumber\\&+&y+z)\phi_T(x)\bigl)\biggl)h_b(x,y,z)+\biggr((1-r^4)(y-z)\phi_A(x) +\nonumber\\&+&r\mu_{0P}\bigr((y-z+(1-r^2)\bar{x})\phi_P(x)+\nonumber\\&+&(-y+z+(1-r^2)\bar{x})\phi_T(x)\bigl) \biggl)h_{q}(x,y,z)\biggl\}
\end{eqnarray}

For decays  with one  vector plus one  pseudo-scalar particles  in the
final state, the analytical formulas are the following:

\begin{itemize}
\item ${B^0 \to D^{*-}M^+}$:
\end{itemize}

\begin{eqnarray}\label{BDK5}
& &  {\mathcal{M}}_{\mathrm{fac}}({{B^0 \to D^{*-}M^+}})  =  \frac{8\pi m_B^4C_F}{\sqrt{2N_C}}\int_0^1   {dx}\,{dy}\,{dz}\phi_B(z)\times\nonumber\\ &\times&\phi_{D}(y)\biggr\{\biggr((1-r^2)y\phi_A(x)-2r\mu_{0P}(1-r^2-yy)\phi_P(x)\biggl)\times\nonumber\\&\times&h_l(x,y,z)-\biggr((1-r^2)(1-(1-r^2)x)\phi_A(x)+\nonumber\\&+&r\mu_{0P}((1-r^2))\phi_P(x)-(1+r^2)\phi_T(x) \biggl)h_{c}(x,y,z)
\biggl\}
\end{eqnarray}
\begin{eqnarray}\label{BDK6}
& & {\mathcal{M}}_{{\mathrm{nfac}}}({{B^0 \to D^{*-}M^+}})   =  \frac{32\pi m_B^4C_F}{\sqrt{2N_C}}\int_0^1   {dx}\,{dy}\,{dz}\,\times\nonumber\\ &\times&\phi_B(z)\phi_{D}(y)\biggr\{-\biggr((1-r^2)((1-r^2)(y+z)+r^2)\phi_A(x)+\nonumber\\&&-r\mu_{0P}\bigr((y+z-(1-r^2)\bar{x})\phi_P(x)+(y+z-2r^2+\nonumber\\ &-&(1-r^2)(1+x))\phi_T(x)\bigl)\biggl)h_b(x,y,z)+\nonumber\\
&+&\biggr((1-r^2)((1-r^2)-r^2(y-z)\bar{x})\phi_A(x) +\nonumber\\ &-&r\mu_{0P}\bigr((y-z-(1-r^2)\bar{x})\phi_P(x)+\nonumber\\&+&(y-z+(1-r^2)\bar{x})\phi_T(x)\bigl) \biggl)h_{q}(x,y,z)\biggl\}
\end{eqnarray}

\begin{itemize}
\item ${B^0 \to D^{*+}M^-}$:
\end{itemize}

\begin{eqnarray}\label{BDK7}
 {\mathcal{M}}_{\mathrm{fac}}({{B^0 \to D^{*+}M^-}})   =  - {\mathcal{M}}_{fac}({{B^0 \to D^{*-}M^+}})
\end{eqnarray}
\begin{eqnarray}\label{BDK8}
& &{\mathcal{M}}_{\mathrm{nfac}}({{B^0 \to D^{*+}M^-}})   =   \frac{32\pi m_B^4C_F}{\sqrt{2N_C}}\int_0^1   {dx}\,{dy}\,{dz} \times\nonumber\\ &&\times\phi_B(z)\,\phi_{D}(y)\,\biggr\{-\biggr((1-r^2)(1-(y+z)r^2-(1-r^2)x) \times\nonumber\\ &&\times\phi_A(x)+r\mu_{0P}\bigr((y+z-(1-r^2)\bar{x})\phi_P(x)-(y+z+\nonumber\\ &&-2r^2-(1-r^2)(1+x))\phi_T(x)\bigl)\biggl)h_b(x,y,z)+\nonumber\\ &&
+\biggr((1-2r^2)(y-z)\bar{x})\phi_A(x) +r\mu_{0P}\bigr((y-z-(1-r^2)\bar{x}) \times\nonumber\\ &&\times\phi_P(x)-(y-z+(1-r^2)\bar{x})\phi_T(x)\bigl) \biggl)h_{q}(x,y,z)\biggl\}
\end{eqnarray}

In the charged decay  modes $B^+\to D_s^{(*)+}\bar K^0$ the analytical
formulas for the  amplitudes are easily obtained from  the ones of the
neutral  modes $B^0\to D_s^{(*)+}  K^-$, given  in Eqs.~(20),(21),(24)
and (25).   The simpler way  to see this,  is that the  operators that
contributes to  the neutral decay modes  are related to  the ones that
contributes to the  charged modes by replacing the  spectator quark of
the $B$  meson $d\to u$  (i.e., $\bar{b}d\to\bar{b}u$), and  the light
quark  from   the  $K$  meson   is  also  replaced  $u\to   d$  (i.e.,
$\bar{u}s\to\bar{d}s$).  The  final result  in the calculation  of the
amplitudes of  the charged modes  are equivalent to the  neutral mode,
with  the replacement  of the  distribution, $\phi_{B^0}\to\phi_{B^+}$
and  $\phi_K\to\phi_{K^0}$.   Also the  analytical  expression of  the
distribution amplitudes are equivalent, with the use of the respective
mass and decay constant of each meson. Besides that, the amplitudes of
decay  modes which  involves two  neutral pseudoscalar  mesons  in the
final  states  are  related  to  the modes  of  charged  pseudoscalars
as                                                 $${\mathcal{A}}(B\to
M_1^0M_2^0)=\frac{1}{\sqrt{2}}{\mathcal{A}}(B\to M_1^+M_2^-)$$

In Eqs.~(\ref{BDK1}) to (\ref{BDK8}) we define the following functions:
\begin{eqnarray}
&&h_c(x,y,z)=D_c(x)\,D_g(x,y)\alpha_s(\mu_f) \bigl(   C_1(\mu_f)  +   {C_2(\mu_f)}/{3}\bigr)\,;\nonumber\\
&&h_{l}(x,y,z)=D_{l}(x)D_g(x,y)\alpha_s(\mu_f)\bigl(   C_1(\mu_f)  +  C_2(\mu_f)/3\bigr)\,;\nonumber\\
&&h_q(x,y,z)=D_q(x)D_g(x,y)\alpha_s(\mu_h) C_2(\mu_h)\,;\nonumber\\
&&h_b(x,y,z)=D_b(x)D_g(x,y)\alpha_s(\mu) C_2(\mu)\,, \label{eq28}
\end{eqnarray}
where    the    functions    $D_b(x,y,z),\,D_d(x,y,z),\,D_u(y)$    and
$D_c=D_c(x)$  and $D_g(x,y)$  are  the quarks  and gluon  propagators,
given by:
\begin{eqnarray}
 D_b^{-1}(x,y,z)&=& m_B^2((y-\bar{z})(1-\bar{x}(1-r^2)))-m_b^2; \nonumber\\ D_q^{-1}(x,y,z)& = & m_B^2\bar{x}(z-y)(1-r^2);\nonumber\\
 D_c^{-1}(x)&=& m_B^2(r^2+\bar{x}(1-r^2))-m_c^2\,;\nonumber\\ D_l^{-1}(y)& = & m_B^2y(1-r^2)\,\nonumber\\
 D_g^{-1}(x,y)&=&k_g^2 = m_B^2 y\,(1-x)\,(1-r^2).
\end{eqnarray}
\noindent  
For  the  charged  modes,  the  Wilson coefficients  that  appear  in
Eqs.~(28)  are exchanged by  $ C_1(\mu_f)/3  + {C_2(\mu_f)}$ for the
first two lines, and by $C_1(\mu)$ for the last two lines.

Note that the perturbative propagators  are going to be substituted by
the non-perturbative ones discussed in the next section.
The scales adopted  in the calculations are the  same as the charmless
case: $\mu=m_b$ for the contribution of the diagram where the gluon is
emitted from the $b$ quark,  and $\mu_h = \sqrt{\Lambda_hm_b}$ for the
contribution where the gluon is emitted from the ``spectator'' quark, with
$\Lambda_h=500$  MeV.   For  the  diagrams  of  Fig.~\ref{figure1}c  and
\ref{figure1}d where the  gluon is emitted from the  quarks in the final
state, we use  the scale $\mu_f=\mu/2=m_b/2$. The QCD  scales are also
the same as the charmless case. 

\section{Regulation of the divergences}

The annihilation amplitudes have  both soft and collinear divergences.
The soft divergences  occur with the vanishing of  the gluon momentum,
and this one has been treated  with an \textsl{ad hoc} cutoff as shown
in Eq.(\ref{eqcut}).  In our work  this procedure is exchanged  by the
introduction  of an IR  finite gluon  propagator obtained  through the
solution        of        Schwinger-Dyson       equations        (SDE)
\cite{cornwall,bp,ap,an2,cornwall2} and consistent with recent lattice
data \cite{abp}. As for the  collinear divergences, we can include the
dynamical mass of  the light quarks, which are  also necessary to keep
the consistency  with the introduction of dynamical  gluon masses. The
poles of the (massive) propagators will give the imaginary part of the
amplitudes.   Therefore,   the  basic  point  in   the  regulation  of
divergences is based on the use of non-perturbative IR information (or
dressed propagators) in the context of the perturbative expansion.  We
will discuss this regulation scheme in detail in the following.

\subsection{IR finite gluon propagator}

A prescription  of how non-perturbative SDE solutions  can be inserted
into the perturbative QCD expansion  was proposed by Pagels and Stokar
many  years ago,  in the  approach denominated  dynamical perturbation
theory (DPT) \cite{pagels}. In their scheme the amplitudes that do not
vanish to  all orders in perturbation  theory are given  by their free
field  values,   while  amplitudes  that   vanish  as  $\lambda\propto
e^{-1/g^2}$ are retained,  and possibly dealt with in  an expansion in
$g^n\lambda$. The work of Ref.\cite{pagels} was particularly concerned
with the  effect of a dynamically  generated quark mass,  but from SDE
\cite{cornwall,bp,ap,an2,cornwall2} and lattice simulations \cite{abp}
we now know that the gluon and coupling constant also have an infrared
finite value,  and the Pagels  and Stokar formulation can  be extended
and  generalized for the  perturbative expansion  using the  quark and
gluon propagators  with a  dynamical mass and  the IR  finite coupling
constant.   Actually   we  expect  that  any   infrared  finite  gluon
propagator  leads to  a  freezing of  the  infrared coupling  constant
\cite{natale2},  meaning that  the  use of  an  infrared finite  gluon
propagator must be accompanied by  an IR finite coupling constant. All
these facts  indicate that we should perform  perturbation theory with
the  dressed quark  and  gluon propagators  and  the effective  charge
(dependent on the gluon mass).

The SDE solution and the lattice result that we shall use were
discussed at length in the following references \cite{abp,trento},
therefore we will not enter into further details about the solutions
and will just present the gluon propagator and coupling constant in
the case that QCD generates a dynamical gluon mass.  We consider a
gluon propagator that will have the form
\be   \imath
{{\Delta}}_{\mu\nu}    (q)=P_{\mu\nu}    {\Delta}(q)+\xi   \frac{q_\mu
  q_\nu}{q^4}  \,\,   ;  \,\,  P_{\mu\nu}=-g_{\mu\nu}   +  \frac{q_\mu
  q_\nu}{q^2} \,\,  , \ee 
where ${\Delta}(q)$  is the gauge  invariant scalar part of  the gluon
propagator,   which   in   Euclidean    space   has   the   form   \be
{\Delta}(Q^2)\propto  \frac{1}{Q^2  + m_g^2  (Q^2)}  \,\,  .  \ee  The
gluonic   SDE  solutions   allow  us   to  write   a   new  propagator
${\hat{\Delta}}^{-1}(Q^2)$ which absorbs all the renormalization group
logs, exactly as happens in  QED with the photon self-energy, and form
the   product  ${\hat{d}}(Q^2)=g^2{\hat{\Delta}}(Q^2)$   which   is  a
renormalization   group   invariant.    The   dynamical   gluon   mass
($m_g^2(Q^2)$) \cite{cornwall}  is given by
\be m_g^2(Q^2)=m_g^2  \left[ \frac{\mathrm{ln} \left( \frac{q^2 + 4
   m_g^2}{\Lambda^2} \right)}{\mathrm{ln} \left( \frac{4 m_g^2}{\Lambda^2}
   \right)} \right]^{- \frac{12}{11}}, 
\ee 
where the   gluon   mass scale $m_g  $  has  typical  values
\cite{cornwall,several}
\begin{equation} m_g = 500 \pm 200\, \mathrm{MeV} \,\,\, .
\end{equation}
As  this  is  a  complicated  expression to  take  into  account  when
performing the  numerical integration, in  practical calculations, the
best we can  do is to assume $m_g^2(Q^2)\approx  m_g^2$.  In this work
the dynamical gluon mass is calculated at the same scales described in
the end of  the Section III.  There is  no significant difference in
the results  if we neglect the  running of the  masses, since slightly
away from the mass value  the effect of momentum dependence takes over
the full propagator effect".

A simple  fit for the coupling  constant that is factored  out in this
procedure is given by  \cite{cornwall} 
\be {\bar{\alpha}}_{sd} (q^2) =
\frac{1}{4\pi b \ln [(4m_g^2 -q^2 -\imath\epsilon )/\Lambda^2]} \,\, ,
\label{eq31}
\ee 
where  $b  = (33-2n_f)/48\pi^2$.   Eq.(\ref{eq31})  clearly shows  the
existence of  the IR  fixed-point as $q^2\rightarrow  0$.  It  must be
stressed that the  fixed point does not depend  on a specific process,
it  is uniquely obtained  as we  fix $\Lambda$  and, in  principle, it
should be  exactly determined  if we knew  how to solve  QCD.  Several
examples of the  use of DPT with the  propagator and coupling constant
discussed above can be found in Ref.\cite{several}.

\subsection{Dynamical quark masses}

The dressed quark propagator can be written as
\be
S(p)=\frac{Z(p^2)}{\imath \gamma \cdot p + M(p^2)} \,\, ,
\label{eqxx}
\ee  where  $Z(p^2)$  is   the  Lorentz  scalar  quark  wave  function
renormalization  and $M(p^2)$ is  the quark  dynamical mass.  At large
momenta, and in the chiral limit,  it is known that that the dynamical
mass is related to the quark condensate ($<{\bar{q}}q>$) as \cite{rob}
\be M(p^2) \propto \frac{<{\bar{q}}q>}{p^2} \,\, .
\label{eqmq}
\ee The  many estimates  of the dynamical  quark mass  ($M(0)$) (among
them  we have  $M(0)\propto <{\bar{q}}q>^{1/3}$)  give  values between
$250\MeV$ and $300\MeV$ \cite{rob,maris}, whereas we can safely assume
$Z(p^2)\approx 1$. The decrease of the mass with the momentum shown in
Eq.(\ref{eqmq}) is not going to be considered, since the mass function
has a plateau at low momenta  and decreases (in the chiral limit) very
fast at  high momenta \cite{maris}, where the  propagator is dominated
by  the  $\imath  \gamma  \cdot  p$  term, in  such  a  way  that  our
calculation  is  not affected  if  we  take  the quark  propagator  as
$S(p)={1}/({\imath \gamma \cdot p + M(0))}$.

\section{Numerical Results}

We perform the numerical  calculation considering the gluon mass scale
with values $m_g  = 400 \MeV, 500 \MeV$.  The  dynamical masses of the
light     quarks    are     $M(0)=m_u=m_d$,     in    the     interval
$250\MeV<M(0)<300\MeV$,  and the  same value  is assumed  for  the $s$
quark, although it has been argued that the constituent $s$ quark mass
may be slightly  higher \cite{maris}. For the the  light mesons, pions
and kaons, we use the asymptotic expression for the wave functions
\begin{eqnarray}
&&\phi^A(x)\,=\frac{f_P}{2\sqrt{2N_C}}6\,x\,(1-x),\qquad \phi^P(x)=\frac{f_P}{2\sqrt{2N_C}} \nonumber\\&& \phi^T(x)=\frac{f_P}{2\sqrt{2N_C}}(2x-1)
\end{eqnarray}
while for the $D(D^*)$  mesons we use
\begin{equation}
\Phi_D(x)\,=\frac{f_D}{2\sqrt{2N_C}}6\,x\,(1-x)\,[1+a_D(1-2x)],
\end{equation}
with   $a_D=0.8$   for   $D$    and   $a_D=0.3$   for   $D_s$   mesons
\cite{Lu:2002iv,Li:2005vu}.  The momentum of  the $B$ meson is carried
almost all by the $b$ quark, and the light quark carries a momentum of
the    order    $\Lambda_{QCD}$.     Providing    that    we    assume
$z\sim\Lambda_{QCD}/m_b$  and   $x,y\gg  z$,  at   leading  order  the
integration over the  $B$ meson wave function is  trivial, yielding to
the decay constant according to  the Eq.~(\ref{normbwave}).  This procedure has been
adopted  in annihilation decay  calculations by  Beneke {\it  et.  al}
\cite{beneke}.

The Wilson's  coefficients are computed  using the equations  given in
the appendices  of Ref. \cite{Lu:2000em}.   We also use  the following
parameters \cite{Amsler:2008zzb}:
\begin{itemize}
 
\item Masses:  

$m_{B}=5.28$  GeV, $m_{B_s}=5.37$ GeV,  $m_D=1.87		$ GeV,  $m_{D_s}=1.97$ GeV, $m_{D^*}=2.01$  GeV, $m_{D_s^*}=2.10$ GeV, $m_b=4.7$
  GeV. $m_c=1.27$ GeV.

\item Lifetimes:

$\tau_{B_d}=1.54$ ps, $\tau_{B_s}=1.466$ ps, $\tau_{B^+}=1.638 $ ps .

\item CKM parameters:

$A=0.814$, $\lambda=0.2257$, $\bar{\rho}=0.135$, $\bar{\eta}=0.349$.
\fussy

\item Decay constants \cite{stone,fdlattice}:

  $f_{B}=190 \MeV$,  $f_{B_s}=236\MeV$ ;   $f_{D}=206\MeV$,
  $f_{D_s}=257\MeV$, $f_{\pi}=130 \MeV$,  $f_K=156\MeV$; 
  $f_{D^*}=245\MeV$,  $f_{D_s^*}=272\MeV$.

\end{itemize}

In  Table   \ref{table1}  we  show   the  results  obtained   and  the
experimental data available for each decay channel.

\begin{table*}[t]
\begin{center}
\caption{\label{table1}Branching  ratios for  $B$  decays.  The  error
  correspond to the variation of the dynamical quark mass, $250 \MeV <
  m_q  < 300  \MeV$. We  show  the results  for the  gluon mass  scale
  $m_g=400\MeV$  and  $500\MeV$. In  the  last  column  are shown  the
  available experimental data.}
\begin{tabular}{p{1.5in}p{1.5in}p{1.5in}p{1in}}           \hline\hline
	
Channel	 & $m_g=400\MeV$ & $m_g=500\MeV$	&    	Experimental Data  \cite{Amsler:2008zzb} \\ [2ex]\hline			
$\bar B^0\to K^+K^-$	& $(4.31 \pm 0.50)\times10^{-8}$ &	$(3.20\pm 0.01)\times10^{-8}$	        & $<4.1\times10^{-7}$ \\[2ex]
$\bar B_s^0\to\pi^+\pi^-$	& $(10.42\pm1.00)\times10^{-7}$ &	$(5.32\pm0.25)\times10^{-7}$  	& $<1.7\times10^{-6}$  \\[2ex]
$\bar B_s^0\to\pi^0\pi^0$	& $(5.21\pm0.05)\times10^{-7}$ &	$(2.66\pm0.13)\times10^{-7}$  	& $<2.1\times10^{-6}$  \\[2ex]
$B^0\to D_s^-K^+$  &	$(6.12\pm0.78)\times10^{-5}$ &	$(3.07\pm0.17)\times10^{-5}$               & $2.9\pm 0.5$ \\[2ex]
$B^0\to D_s^+K^-$  &	$(6.15\pm0.71)\times10^{-8}$ &	$(2.92\pm0.23)\times10^{-8}$              	&  \\[2ex]
$B^+\to D_s^+\bar{K^0}$ &	$(0.89\pm0.16)\times10^{-8}$ &	$(0.32\pm0.11)\times10^{-8}$   	& $<9\times10^{-4}$ \\[2ex]
$B_s^0\to D^-\pi^+$  &	$(1.72\pm0.04)\times10^{-6}$ &	$(1.18\pm0.40)\times10^{-6}$	&   \\[2ex]
$B_s^0 \to \bar{D}^0\pi^0$  &	$(0.86\pm0.02)\times10^{-6}$ &	$(0.61\pm0.17)\times10^{-6}$	&    \\[2ex]
$B_s^0\to D^+\pi^-$  &	$(5.80\pm0.60)\times10^{-7}$ &	$(2.51\pm0.96)\times10^{-7}$	&    \\[2ex]
$B_s^0\to D^0\pi^0$ &	$(2.90\pm0.30)\times10^{-7}$ &	$(1.26\pm0.48)\times10^{-7}$	&   \\[2ex]
$B^0\to D_s^{*-}K^+$ &	$(2.85\pm0.18)\times10^{-5}$ &	$(1.90\pm 0.06)\times10^{-5}$ & $ 2.2\pm 0.6$ \\[2ex]
$B^0 \to D_s^{*+}K^-$ &	$(6.23\pm0.50)\times10^{-8}$ &	$(2.68\pm0.07)\times10^{-8}$ &  \\[2ex]
$B^+\to D_s^{*+}\bar{K^0}$  &	$(1.71\pm0.52)\times10^{-8}$ &	$(1.28\pm0.22)\times10^{-8}$ 	&  $<9\times10^{-4}$ \\[2ex]
$B_s^0\to D^{*-}\pi^+$   &	$(3.36\pm0.06)\times10^{-6}$ &	$(1.88\pm0.21)\times10^{-6}$	& \\[2ex]
$B_s^0\to D^{*+}\pi^-$   &	$(3.26\pm0.29)\times10^{-8}$ &	$(3.92\pm1.31)\times10^{-8}$	& \\[2ex]
\hline
\end{tabular}
\end{center}
\end{table*}
\section{Conclusion}

We have  studied some  decay channels of  $B^0_s,\,B^0_d,\,B^+$ mesons
which  occur through the  annihilation diagrams.  We have  argued that
infrared  finite  gluon  propagators  and running  coupling  constants
obtained as solutions of  the QCD Schwinger-Dyson equations, may serve
as a natural cutoff for  the end-point divergences that appear in the
calculation of these decays.  Also,  we use the constituent quark mass
for the  light quarks  in the  propagators to be  able to  perform the
twist-3 calculation of the amplitudes.

Comparing   to  the   data  available,   our  calculations   are  very
satisfactory, notably  for the  two channels with  actual measurements
$B^0\to D_s^-K^+$  and $B^0\to  D_s^{-*}{K}^+$.  The results  are also
reasonably compatible  with some predictions  of the pQCD in  the $k_T$
factorization,               see              for              example
Refs.~\cite{Chen:2000ih,Lu:2002iv,Li:2003wg,Li:2005vu,Ali:2007ff,Li:2008ts,Zou:2009zza}.

We can see  from the results that the variation  on the dynamical mass
of the quarks generally introduces only a small error.  The numerical
results are  more dependent on the gluon  mass scale, as we  can see a
more  significant  variation of  the  results  with  this mass  scale.
Comparing to the available data  it is possible to narrow the interval
for the  gluon mass  scale for values  closer to  $m_g=500\MeV$.  This
value  for  the  gluon  mass  scale is  in  accordance  with  previous
calculations \cite{several}.  Future  measurements for the other decay
modes,  besides  $B^0\to  D_s^-K^+$  and $B^0\to  D_s^{-*}{K}^+$,  can
establish a  more precise value  for the $m_g$ parameter.  Besides the
annihilation  modes,  it  is   also  possible  to  apply  the  present
regulation prescriptions to other types of $B$ decay.

The already existent agreement of  some B decays with the experimental
data and  the possible  future confirmation of  the other  rare decays
computed   here,   may  indicate   that   the   introduction  of   the
nonperturbative information provided  by the Schwinger-Dyson equations
in the form of dynamical masses for quarks and gluons, consistent with
the most  recent lattice QCD  simulations, is gathering more  and more
robust  phenomenological  results   showing  that  these  mass  scales
possibly cannot  be neglected when performing  high precision hadronic
calculations.

\section{Acknowledgments}
 This research was supported by the FAPESP (CMZ) and CNPq (AAN).

\end{document}